\documentclass[a4paper,11pt]{article}
\pdfoutput=1 

\usepackage[table]{xcolor}
\usepackage{tcolorbox} 
\definecolor{mygray}{gray}{0.6}
\usepackage{jheppub} 
            
\usepackage{bm,amssymb,slashed,graphicx,multirow,soul,mathtools,xspace,array,comment}  
\usepackage{float}                    
\allowdisplaybreaks 
\usepackage{ bbold }
\usepackage{color}
\usepackage{subfigure}
\usepackage[printonlyused]{acronym}
\usepackage{hyperref}
\usepackage{physics}

\usepackage{overpic}

\acrodef{PDG}[PDG]{Particle Data Group}
\acrodef{OPE}[OPE]{Operator Product Expansion}
\acrodef{FCNC}[FCNC]{flavour-changing neutral current}
\acrodef{RHC}[RHC]{right-handed currents}
\acrodef{SM}[SM]{Standard Model}
\acrodef{NP}[NP]{New Physics}
\acrodef{MFV}[MFV]{Minimal Flavour Violation}
\acrodef{SD}[SD]{short-distance}
\acrodef{LD}[LD]{long-distance}
\acrodef{DA}[DA]{distribution amplitude}

\newcommand{\VEV}[1]{\langle #1 \rangle} 
\newcommand{\state}[1]{|#1\rangle}

\newcommand{\matel}[3]{\langle #1|#2|#3\rangle}
\newcommand{\al}{\alpha}
\newcommand{\sig}{\sigma}
\newcommand{\be}{\beta}
\newcommand{\ga}{\gamma}
\newcommand{\de}{\delta}
\newcommand{\De}{\Delta}
\newcommand{\la}{\lambda}
\newcommand{\eps}{\epsilon}

\newcommand{\GeV}{\,\mbox{GeV}}
\newcommand{\MeV}{\,\mbox{MeV}}
\newcommand{\ORD}{{\cal O}}

\newcommand{\TAB}{Tab.~}
\newcommand{\FIG}{Fig.~}

\newcommand{\SEC}{Sec.~}
\newcommand{\SECs}{Secs.~}
\newcommand{\APP}{App.~}
\newcommand{\APPs}{Apps.~}
\newcommand{\EQ}{Eq.~}

\newcommand{\mB}{m_{B}}

\newcommand{\mDs}{m_{D_s}}

\newcommand{\mBs}{m_{B_s}}

\newcommand{\lscale}[1]{{ |#1\!\GeV}}
\newcommand{\qbarq}{\langle \bar{q}q\rangle}
\newcommand{\sbars}{\langle \bar{s}s\rangle}

\newcommand{\ZB}{Z_B}

\newcommand{\pBsq}{p^2}
\newcommand{\pBpsq}{\tilde{p}^2}
\newcommand{\spr}{ \tilde{s} }

\newcommand{\mplus}{m_+}

\newcommand{\MSbar}{\overline{\text{MS}}}

\definecolor{violet}{rgb}{0.94, 0.2, 0.8}
\definecolor{lightblue}{rgb}{0.39, 0.58, 1.00} 
\definecolor{lightgreen}{rgb}{0.1, 0.73, 0.33}

\numberwithin{equation}{section}

\newcommand*{\mathcolor}{}
\def\mathcolor#1#{\mathcoloraux{#1}}
\newcommand*{\mathcoloraux}[3]{%
  \protect\leavevmode
  \begingroup
    \color#1{#2}#3%
  \endgroup
}

\title{ \boldmath  Isospin Mass Differences of the $B$, $D$ and $K$ }
\author[1]{Matthew Rowe,}
\author[1,2]{Roman Zwicky}

 \affiliation[1]{Higgs Centre for Theoretical Physics, School of Physics and Astronomy, University of Edinburgh, \\Edinburgh EH9 3JZ, Scotland}
\affiliation[2]{ Theoretical Physics Department, CERN,
Esplanade des Particules 1, \\ Geneva CH-1211, Switzerland}

\emailAdd{m.j.rowe@sms.ed.ac.uk}
\emailAdd{roman.zwicky@ed.ac.uk}

\abstract{We compute the electromagnetic mass difference for the $B$-, $D$- and $K$-mesons 
using QCD sum rules with double dispersion relations.  
For the $B$- and $D$-mesons we also compute the linear quark mass correction, 
whereas for the $K$ the standard soft theorems prove more powerful. 
The mass differences, which have not previously been computed via a double dispersion, are fully consistent with experiment, albeit with large uncertainties.
}

\begin{document}
\preprint{CERN-TH-2023-005}

\toccontinuoustrue

\maketitle

\flushbottom

\setcounter{tocdepth}{3}
\setcounter{page}{1}
\pagestyle{plain}


\section{Introduction}
\label{sec:intro}

The mass difference of charged and neutral hadrons,
\begin{equation}
\label{eq:Delta}
\Delta m_H  = m_{H^+} - m_{H^0}  \;,  \qquad H = B,D,K,\pi, p \;,
\end{equation}
is an isospin breaking effect and 
has intrigued particle physicists from the very beginning.   
In particular  the proton-neutron \cite{Zee:1971df}
and the $\pi^+$-$\pi^0$ \cite{Das:1967it} mass difference have been discussed extensively. 
At the microscopic level $\Delta m_H$  is driven by differences in  the electric charge and  the mass $m_q$ 
of the hadron's  light valence  quark $q =u,d$
\begin{equation}
\label{eq:main}
\Delta m_B =  \Delta m_B |_{\text{\text{QED}}}  + \Delta m_B |_{m_q}  \;.
\end{equation} 
The sign and the size depends  on the hadron in question 
and QED stands for quantum electrodynamics.\footnote{Strictly speaking the separation \eqref{eq:main} is not well-defined as it requires 
fixing a (quark mass) renormalisation scheme e.g.  \cite{Borsanyi:2014jba}.  
In turn this is a reason for being interested in the problem as, especially light, quark masses cannot be determined to high precision without folding in QED. 
This shows for example in the $D$-meson results in comparison between  \cite{Borsanyi:2014jba} and \cite{Giusti:2017dmp}.
For our purposes $ \Delta m_B |_{m_q} $ is as defined from \eqref{eq:DelmBmq}.}$^,$\footnote{Effects due 
to the weak force are of $\ORD(\Lambda^2_{\text{QCD}}/m_W^2)$ with respect to QED and are thus negligible. 
Similar effects are relevant in the context of neutral meson mixing  e.g 
\cite{Bigi:2000yz,Branco:1999fs}.}
Recent  lattice Monte Carlo simulations  \cite{Borsanyi:2014jba,Giusti:2017dmp}
have verified this to a high accuracy, for light and charm mesons, by computing both the charged and the neutral  mass and effectively using \eqref{eq:Delta}.

One may take a different approach and compute the two differences  in \eqref{eq:main} separately
by using the second order perturbation theory formula
(with $H = B$ for definiteness)\footnote{Note that in the literature  the notation 
 $\Delta m_B^2 \equiv 2 m_B \Delta m_B$ is also frequently used.} 
\begin{equation} 
\label{eq:DelmBQED}
\de m_B |_{\text{QED}}  =  
 \frac{- i  \al}{2 m_B (2 \pi)^3}    \int d^4 q \, 
 T^{(B)}_{\mu\nu}(q) \Delta^{\mu \nu}(q) + \ORD(\al^2) \;, 
\end{equation}
with
\begin{equation}
\De m_B |_{\text{QED}} \equiv  \de m_{B^+} |_{\text{QED}} - \de m_{B^0} |_{\text{QED}} \;,
\end{equation}
known  in the  current algebra era \cite{Feynman:1954xxl,Cini:1959szx}.
Above $ \Delta_{\mu \nu}(q) = \frac{1}{q^2}( - g_{\mu\nu} + (1-\xi) \frac{q_\mu q_\nu}{q^2})$ is the  photon 
propagator, $\al = e^2/(4 \pi)$  the fine structure constant   and  $T^{(B)}_{\mu\nu}(q)$ is the (uncontracted) forward Compton scattering tensor, 
 \begin{equation}
\label{eq:Tmunu}
T^{(B)}_{\mu\nu}(q) = i   \int d^4 x e^{-i q\cdot x}     \matel{B}{T j_{ \mu} (x) j_{\nu}(0)}{B} \;,
\end{equation}
with $j_\al = \sum_q Q_q \bar q \ga_\al q$, the electromagnetic current.

In 1963, Cottingham \cite{Cottingham:1963zz} improved this formula by parameterising it in terms of form factors and relating 
it to structure functions. That is, by deforming the contour $q_0 \to i q_0$ and writing a 
dispersion representation, assessing the number of subtraction terms
of the form factors thus allowing him to write the contribution as an integral over $Q^2 = -q^2 \geq 0$ and 
$\nu  = p \cdot q/m_B$ in the physical region.   
This opened the gate for many phenomenological 
studies saturating the dispersion relation by a few terms beyond the elastic one and using high energy constraints.  
This is a formidable task as one requires the knowledge of a correlation function over the entire energy 
range akin to the situation of the vacuum polarisation for the anomalous magnetic moment.
Some examples are for $K$, $\pi$ \cite{Donoghue:1996zn,Bardeen:1988zw}  using chiral perturbation theory 
 (and large $N_c$), 
 for $B$ and $D$  \cite{Colangelo:1997tc,Luty:1995zx} using heavy quark theory (and large $N_c$), 
 for the proton-neutron \cite{Walker-Loud:2012ift} with updated fits to the structure functions  
 and an approach to $B$, $D$, $K$ and $\pi$ using vector meson dominance \cite{Hambye:1993gr}.
 Another interesting point, not unrelated, is that \eqref{eq:DelmBQED} requires renormalisation  \cite{Collins:1978hi}  and 
it was argued that it is justified to cut-off  the $Q^2$-integral.  
Debates about subtraction terms are ongoing cf.  \cite{Walker-Loud:2012ift}  and the response  \cite{Gasser:2015dwa}.

Here we do \emph{not} follow this phenomenological approach  but evaluate \eqref{eq:Tmunu} directly in Minkowski space  using double dispersion relation sum rules 
and thus determine the mass differences from a unified framework (i.e. same hadronic input).\footnote{This function has been evaluated for the 
pion on the lattice with good agreement with experiment only very recently using the infinite volume 
reconstruction method \cite{Feng:2021zek}.} 
To the best of our knowledge this has not been done previously with sum rules, presumably  due to   the subtleties of non gauge-invariant interpolating currents \cite{Zwicky:2021olr,Nabeebaccus:2022jhu}.
For example, in leptonic decays this requires the introduction of a non-local 
interpolating operator (or an auxiliary scalar field carrying the charge to infinity) for gauge invariance 
and reproduction of all infrared sensitive logs \cite{Nabeebaccus:2022jhu}.
However, in the case at hand this is not necessary, as verified
by explicit computation,  since $\Delta m_B $ is an infrared safe quantity. 

An efficient and transparent  way to implement the first order quark mass corrections is to make use of 
the Feynman-Helmann theorem which gives 
\begin{equation}
\label{eq:FH}
 m_B^2|_{m_q} = \sum_q  m_q  \matel{{B}}{\bar qq}{{B}}  \;,
\end{equation}
as rederived in \APP\ref{app:FH}. For the difference \eqref{eq:Delta} this gives
\begin{equation}
\label{eq:DelmBmq}
\Delta m_B \left|_{m_q}  \right.  =   \frac{(m_u- m_d)}{2 m_B}   \matel{{B}}{\bar qq}{{B}}  + \ORD((m_u- m_d)^2) \;.
\end{equation}
The matrix element $ \matel{{B}}{\bar qq}{{B}} $ can be evaluated in the isospin degenerate limit $q = u = d$
since we work to leading order (LO).  For the $B$- and the $D$-meson we compute this matrix element 
whereas for the Kaon and the pion a soft theorem 
$ \matel{{\pi}}{\bar qq}{{\pi}} = -  \frac{2}{f_\pi^2}  \vev{\bar qq} + \ORD(m_\pi^2/m_\rho^2)$, 
with $f_\pi \approx 131 \MeV$),  due to their pseudo-Goldstone nature, proves  more effective.

In principle one could compute all the $\Delta m_B |_{m_q}$-effects 
with  the QCD analogue of \eqref{eq:DelmBQED} but this would be rather inefficient and we further comment 
in the relevant section. Another noteworthy aspect is that we were not able to obtain stable sum rules for the pion (cf. \SEC\ref{sec:KQED}).

The paper is organised as follows. In \SEC\ref{sec:QED} the electromagnetic computation is presented, 
followed by the quark mass correction in \SEC\ref{sec:mq}. 
We give an overview of the results and the conclusions in \SEC\ref{sec:conclusions}. 
Comments on quark hadron duality, the numerical 
input. some (extra) computation  and  useful classic results 
are collected in  \APPs \ref{app:duality}, \ref{app:self}, \ref{app:input} and \ref{app:classic}
respectively.

\section{Electromagnetic Mass Difference $\Delta m_H|_{\text{QED}}$ from QCD Sum Rules}
\label{sec:QED}

The electromagnetic mass difference follows from the formula quoted in \eqref{eq:DelmBQED}
and it is our task to evaluate this. The main theoretical challenge is to incorporate 
the two hadrons for which a non-perturbative method is needed. We use QCD sum rules  \cite{SVZ79I}
with a double dispersion relation.  The first step involves the adaption of an interpolating operator. 
For the heavy mesons a pseudoscalar current is suitable and has proven to give good results in many 
other contexts.  For particles of light quark masses, and Goldstone particles in particular 
\cite{Novikov:1981xi}, pseudoscalar interpolating operators are unsuitable as they are 
infested by so-called direct instantons \cite{Shuryak:1982qx}.\footnote{For the heavy mesons 
axial interpolating operators are unsuitable because the $1^+$ states are relatively low, e.g. 
for the $J^P = 0^-$ $B$-meson with $m_B \approx 5.28\GeV$ there is a $1^+$ $B_1(5721)$ 
with $m_{B_1} \approx 5.72 \GeV$.  This is too close to the two pion threshold and even below 
the typical continuum threshold $s_0 \approx (6 \GeV)^2$  assumed for the pseudoscalar 
operators.}  We therefore discuss the heavy mesons and the $K$-meson separately in 
\SECs \ref{sec:BDQED} and \ref{sec:KQED} respectively. 

An important criteria in assessing the validity of our sum rules is the so-called daughter sum rule 
which we consider worthwhile to present now. 
In the simple single dispersion relation case this criteria reads
\begin{eqnarray}
\label{eq:DSR}
m_B^2(s_0,M^2) =    \int_{\textrm{cut}}^{s_0} e^{-s/M^2} \rho(s) s ds  / ( \int_{\textrm{cut}}^{s_0}  e^{-s/M^2} \rho(s)    ds)  \;,
\end{eqnarray}
where $M^2$ is the Borel parameter, the ``cut" marks the onset of physical states, 
 $\rho(s) = r_B \de(s- m_B^2)  + \dots $ is the spectral density 
and the dots stand  for states above the continuum threshold $s_0$.
Formally, the residue $r_B$ drops out in the ratio. In practice $\rho(s)$ is a continuous function 
in partonic computations
and  \EQ\eqref{eq:DSR} should 
be seen as a self-consistency criteria for an $s_0$ in the range of $(m_B + 2 m_\pi)^2$ of 
 $(m_B + 4 m_\pi)^2$.  If that is the case then  \EQ\eqref{eq:DSR}   can be used to 
 fix the central value of $s_0$.

\begin{figure}[ht]
 \centerline{
\begin{overpic}[width=5.0in]{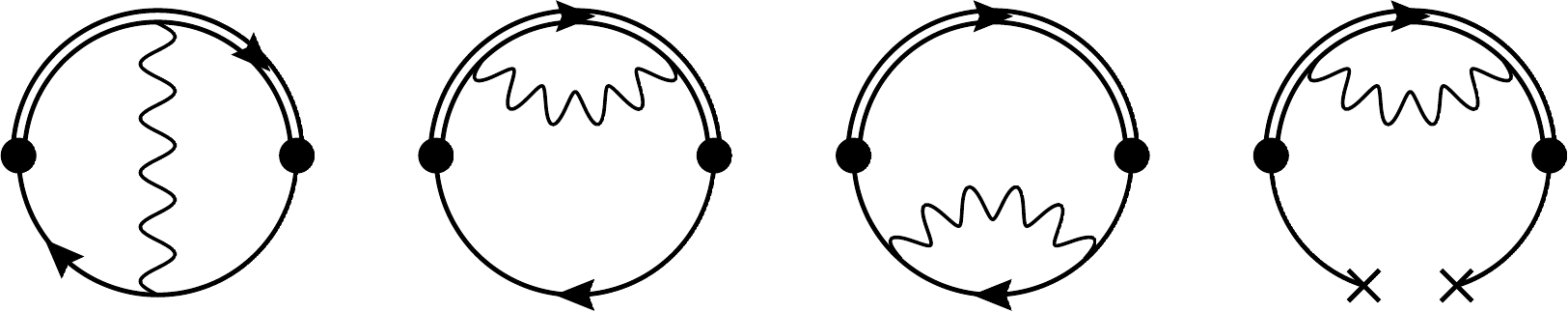}
\put(4,13){$b$}
\put(4,5){$\bar q$}
\put(12,10){$\gamma$}
\end{overpic}}
 \caption{\small  Diagrams contributing to the correlation function in \eqref{eq:pseudo} 
 with the double line representing the $b$-quark. 
 (left) main diagram of the $Q_b Q_q$ mixed type. (middle) $b$- and $q$-quark self energies. 
 (right) $\VEV{\bar qq}$-condensate part to $b$-quark self energy. 
 There is no corresponding part for the $q$-quark self energy since $\VEV{\bar bb}$ is negligibly small. 
 For the mass difference only 
 the first one is relevant while the others are useful to obtain stable sum rules as described in the text.}
 \label{fig:diaQED}
 \end{figure}

\subsection{$B$- and $D$-meson with Pseudoscalar Operators}
\label{sec:BDQED}

As motivated at the beginning of the section, the default choice for 
 heavy-light $0^-$ meson interpolating operators are
\begin{equation}
\label{eq:JB} 
 J_B  = m_+ \bar b i \ga_5 q  \;, \quad 
\ZB  \equiv \matel{\bar B }{J_B}{0 } = m_B^2 f_B  \;, \quad m_+  \equiv (m_b + m_{q}) \;.
\end{equation}
  In determining \eqref{eq:DelmBQED}, one of the main challenges, is 
that the momenta for the two $B$-meson is degenerate. We  bypass this 
problem by introducing an auxiliary momentum $r$ into one of the currents and let it flow out at one of 
the two interpolating operators. 
Concretely we start from 
\begin{alignat}{2}
\label{eq:pseudo}
& \Gamma_{qq'}(\pBsq,\pBpsq) &\;=\;& c  \, i^3  \int_{x,y,z,q}  e^{i ( \tilde{p}z-ip y-(q + r)x)} 
\matel{0}{T J_B^\dagger(z) j_{\mu}(x)j_{\nu}(0) J_B(y)}{0}\Delta^{\mu\nu}(q)|_{Q_q Q_{q'}}
  \nonumber \\[0.1cm]
& &\;=\;&  \int_0^\infty  ds  \int_0^\infty d\spr \frac{  \rho_{\Gamma_{qq'}}(s,\spr) }{( s- \pBsq) 
( \spr-\pBpsq) } = \frac{Z_B^2 \de_{qq'} m_B }{( m_B^2- \pBsq) 
( m_B^2 -\pBpsq) } + \dots \;,
\end{alignat}
with  $ c \equiv  \frac{- i  \al}{2 m_B (2 \pi)^3} $, $\tilde p = p+r$, 
shorthands
$xp = x \cdot p$,  $\int_{q,x}  = \int d^4 q d^4 x$  and the density is given by 
\begin{equation}
\label{eq:disc}
( 2\pi i)^2  \rho_{\Gamma_{qq'}}(s,\spr) =  \text{disc}_{s,\spr} [\Gamma_{qq'}(s,\spr) ] \;,
\end{equation}
the double discontinuity with further relevant explanations at the end of the section. The quantity  $\Delta_{qq'} m_B$ denotes the part proportional 
to the $Q_q Q_{q'}$-charges.   Of course the auxiliary momentum  $r$ has to disappear from the final result. 
This is achieved by the on-shell condition ``$\pBpsq = \pBsq$" 
and is implemented in practice by treating them equally ($p$-$\tilde p$ symmetry) and requiring the daughter sum rule to be satisfied reasonably well.
The QCD sum rule is then given by
\begin{equation}
\de_{qq'} m_B   = \frac{1}{Z_B^2} \int_{m_+^2}^{\bar{\de}^{(a)}(m_+^2)} ds\, e^{\frac{(m_B^2-s)}{M^2}}   \int_{m_+^2}^{\bar{\de}^{(a)}(s)}  d\spr  \, e^{\frac{(m_B^2-\spr)}{M^2}} 
 \rho_{\Gamma_{qq'}}(s,\spr)  \;,
\end{equation}
where $M^2$ is the Borel parameter from the Borel transformation and the $\bar{\de}^{(a)}$ 
is the continuum threshold
\begin{equation}
\label{eq:de}
\bar{\de}^{(a)}(s) =  2^{1/a} \sigma_0 \left(1 - \left(\frac{ s}{ 2^{1/a} \sigma_0} \right)^{a} \right)^{1/a} \;,
\end{equation}
which is complicated for double dispersion sum rules \cite{Balitsky:1988tpa}. Here it is implemented 
as in \cite{Pullin:2021ebn} but simplified since the two hadrons are identical implying $M^2 \rightarrow 2 \hat{M}^2$ and $\tilde{s}_0 = \tilde{t}_0 = \sigma^{(a)}_0 2^{1/a}$ (allowing for  elimination of those parameters). 
The number  $ \sigma_0 \approx 35 \GeV^2$  takes on the r\^ole of  $s_0$ in \eqref{eq:DSR} 
and we shall use the notation $s_0 \equiv \sig_0$ hereafter for reasons of familiarity. 
The parameter $a$ is a model-parameter and the independence of the result is a measure of the quality 
of the result itself. 

Let us turn to the computation. 
In perturbation theory there is the diagram connecting the $q$- to the $b$-quark and the self energies. 
We focus on the former, as it is numerically dominant, and present the self energies and the condensate contribution in \APP\ref{app:self}.  The computation can be done analytically and we obtain the following compact result for the density 
\begin{equation}
\label{eq:rhoPbq}
\rho_{\Gamma_{bq}} =  \frac{  N_c  \al Q_q Q_b m_+^2 }{32 \pi^3 m_B}  \cdot \frac{ \sqrt{\la \tilde{\la}}
 }{s \spr} \bigg(   A + \frac{B}{\mathtt{b}} \ln \Big( \frac{\mathtt{a}+\mathtt{b}}{\mathtt{a}-\mathtt{b}} \Big) \bigg) \;,
\end{equation}
where 
\begin{alignat}{2}
& \mathtt{a} &\,=\,& m_q^2 - \frac{1}{4 \sqrt{s \spr} } \big( s \spr + ( m_+ m_-)^2 \big)  + \Big\{ q \leftrightarrow b \Big\} \;,\quad 
\mathtt{b} = \frac{1}{2} \sqrt{ \frac{ \lambda \tilde{\lambda}}{ s \spr  }}   \;,\quad 
A =  m_-^2 \;,  \nonumber \\[0.1cm]
& B &\,=\,&  \Big\{  Y \tilde{Y} s \spr + \frac{1}{2}m_q^2  \sqrt{s \spr} (Y + \tilde{Y}) - \frac{1}{4} m_-^2 \Big( s + \spr + 4m_bm_q +2m_q^2 \Big) - \frac{1}{4}\mplus^2 \sqrt{s \spr} \Big\}  + \Big\{ q \leftrightarrow b \Big\} \;, \nonumber
\end{alignat}
with further abbreviations
\begin{equation}
m_\pm = m_b \pm m_q \;, \quad \lambda = \lambda(s,m_b^2,m_q^2) \;, \quad 
Y = \frac{s- m_+ m_-}{2s} \;,
\end{equation}
$\la(x,y,z)= x^2 + y^2 +z^2 - 2xy - 2xz -2 yz$ is the K\"all\'en function and in the tilde quantities  
 $\tilde{Y}$ and $\tilde{\la}$ 
 we have $s \to \spr$.  
 
 A few words about the computation. We have taken the discontinuity in \eqref{eq:disc} using  
 Cutkosky rules. A crucial point is that we do not cut the photon propagator as this would be a 
 QED correction to the $B$-meson state and does not contribute to \eqref{eq:DelmBQED}.
 This amends the meaning of  \eqref{eq:disc}.  
 
 Let us turn to the usage of the auxiliary momentum $r$ in the context of double dispersion sum rules. 
  First we note that this is different to 
 a form factor computation, e.g. $F^{\pi \to \pi}(q^2)$ \cite{Nesterenko:1982gc}, where the momentum transfer 
 naturally takes on the r\^ole of this variable.  It is closer to $\Delta F = 2$ matrix elements as there is no momentum transfer but the flavour contractions naturally lead to a symmetric configuration (e.g. 
 \cite{Kirk:2017juj})  which is more straightforward. 
In fact since our procedure \eqref{eq:pseudo} artificially
breaks the $bq$-symmetry, $\mathtt{a}$ and $B$ turn out to be non-symmetric whereas 
 $\mathtt{b}$ and $A$ remain symmetric. This has to be  remedied by  the following substitution 
\begin{equation}
\mathtt{a} \to \frac{1}{2} ( \mathtt{a} + \mathtt{a}|_{b\, \leftrightarrow\, q}) \;, \quad B \to \frac{1}{2} ( B + B|_{b\, \leftrightarrow\, q}) \;,
\end{equation}
which is apparent from the way the Cutkosky cuts work out. 
We have performed the computation in general gauge. 
Of course $\Gamma_{qq'}$ is gauge dependent 
but as stated earlier its discontinuity in the $bq$-quark lines are not. 
This is the case since the particles are put on the mass shell and it is important that the quantity is infrared safe.  Otherwise, as previously stated, one needs to introduce extra machinery \cite{Nabeebaccus:2022jhu}.

\subsubsection{Numerics}
\label{sec:numerics}

Our numerics have three cornerstones, the hadronic input parameters in \TAB\ref{tab:inputParams}, the daughter sum rule \eqref{eq:DSR} 
and the choice of a mass scheme for $m_b$. Whereas there is nothing to say about point one, the others are in need of some explanation. 
We start with the $B$-meson case. 
The daughter sum rule constrains the sum rule parameters: the continuum threshold $s_0$ and the Borel parameter $M^2$.  
Additional constraints, defining the Borel window, are  the convergence of the condensate expansion and keeping the $B$-pole term dominant 
versus the continuum contribution \cite{SVZ79I}. Let us turn to the question of the mass scheme which is not independent of the second point. 
We consider  the pole-, the kinetic- and the $\MSbar$-scheme. 
In the pole scheme the $b,c$-quark self energy contributions (perturbative and condensate, diagrams 2 and 4 in \FIG\ref{fig:diaQED}) vanish and the sum rules are not stable, that is no Borel window, and we therefore discard it. For the $\MSbar$-scheme  the $b$-quark self energies are dominant with the $b$-$q$ contribution comparable to the condensates.  Since these contributions cancel in the observable $\Delta m$, this scheme 
is not ideal either and we therefore drop it. Hence we are left with the kinetic scheme for the $b$-quark 
which shows good properties as for the $B \to \ga$ form factor \cite{Janowski:2021yvz} 
and the $g_{BB*\ga}$-couplings \cite{Pullin:2021ebn}. For the $c$-quark the self energies 
are not dominant  and we use the $\MSbar$-scheme, also because the kinetic-scheme has proven unsuitable in for $g_{DD*\ga}$ \cite{Pullin:2021ebn}.

As stated above the daughter sum rule \eqref{eq:DSR} is used to fix $s_0$. 
For that purpose it  is instructive to define the normalised ratio
 \begin{equation}
 \label{eq:U}
  U(s_0,M^2) \equiv      \frac{1}{m_B^2} \cdot   m_B^2(s_0,M^2) \;,
 \end{equation}
 of the sum rule value over the experimental one which has to be close to unity for self-consistency of
 the approach. This leads to 
\begin{equation}
\label{eq:axial}
\{  s_0, \hat{M}^2 \}_{B}   = \{  35.2(1.0), 2.6(0.5)\}  \GeV^2  \;, \quad \{  s_0, \hat{M}^2 \}_{D}   = \{  5.5(1), 1.0(0.25)\}  \GeV^2   \;,
\end{equation}
for which  
 \begin{equation*}
 U(s_0 \pm 1 \GeV^2 ,M^2)_{\Delta m_B|_{\text{QED}} } =  1 \pm 0.01   \;, \quad  
 U(s_0 \pm 0.1 \GeV^2 ,M^2)_{\Delta m_D|_{\text{QED}} } = 1 \pm 0.01 \;.
 \end{equation*}

Using the input parameters in \TAB\ref{tab:inputParams} 
(with $m_b^{kin}(1\GeV),\bar{m}_c(\bar{m}_c)$) and the $f_{B,D}$ sum rule to LO (cf. \APP\ref{app:decayConst})
for the $Z_B$-factor we get
\begin{equation}
\label{eq:mHQED}
\Delta m_B|_{\text{QED}} =  +1.58^{+0.26}_{-0.23} \MeV \;, \quad  \Delta m_D|_{\text{QED}} =  +2.25^{+0.89}_{-0.52} \MeV \;,
\end{equation}
where the error is obtained by adding the individual errors in quadrature. The dominant error is due to the heavy quark mass $m_{b (c)}$ ($50$-$60\%$). The Borel mass $M^2$ and duality parameters $a$ each contribute a $20$-$25\%$ uncertainty. The error in $a$ is quantified by taking the standard deviation of the results with $a \in [\frac{1}{2},1,2,\infty]$. The errors for the $D$-meson are larger reflecting the generically inferior quality of the sum rule.

\subsection{$K$-meson with Axial Operators}
\label{sec:KQED}

As explained at the beginning of this section pseudo Goldstone bosons cannot 
be interpolated by pseudoscalar operators and one therefore resorts to axial ones 
\begin{equation}
\label{eq:A}
A_\mu =  \bar q  \, \gamma_{\mu} \gamma_{5} \, s \;,    \quad 
\matel{0} { A_\mu }{ K(p)}  = i  p_{\mu} f_{K} \;.
\end{equation}
The  correlation function corresponding to \eqref{eq:pseudo} assumes the form
\begin{alignat}{2}
\label{eq:axial}
& \Gamma^{\al\be}_{qq'}(\pBsq,\pBpsq) &\;=\;& c   i^3 \int_q \int_{x,y,z}  e^{i ( \tilde{p}z- p y-(q + r)x)} 
\matel{0}{T A^{\al}(z) j_{\mu}(x)j_{\nu}(0) A^{\dagger \, \be}(y)}{0} \Delta^{\mu\nu}(q)|_{Q_q Q_q'}
  \nonumber \\[0.1cm]
& &\;=\;&   g_{\al \be}  {\Gamma}^{(0)}_{qq'}  + p_{\al}  p_{\be} {\Gamma}^{(2)}_{qq'}  + \ORD(r)  \dots \;,
\end{alignat}
where the $\ORD(r)$-terms are not of interest to us. The decisive information is in the $p_\al p_\be$-term 
which takes on the form 
\begin{equation}
\label{eq:F2tilde}
{\Gamma}^{(2)}_{qq'}  =  \frac{f_K^2 \de_{qq'} m }{( m_K^2- \pBsq) 
( m_K^2 -\pBpsq) } + \dots \;, 
\end{equation}
in a hadronic representation where the dots represent higher states in the spectrum (which includes the $K^*$-meson in this case). 

Let us turn to the computation which involves some practical matters. Computing the double discontinuity 
of ${\Gamma}^{(2)}_{qq'}$ is laborious as there are open Lorentz indices. One may though obtain the same information from a linear combination of \eqref{eq:pseudo} and  
\eqref{eq:axial} with contracted indices. It follows from Ward identities that  ($ d= 4$)
\begin{equation}
{\Gamma}^{(2)}(s,s)  = \frac{1}{s^2(1-d)}\left( s  \Gamma^{\al}_{\al}(s,s) - d\, \Gamma(s,s) )   \right) \;,
\end{equation}
where we omitted the $qq'$-subscript for brevity and have set $ s= \spr$. 
The generalisation to the  $ s \neq \spr$ is in principle ambiguous but fortunately the differences 
are not that sizeable. Concretely we use
\begin{equation}
\label{eq:pres}
{\Gamma}^{(2)}(s,\spr)  =
\frac{1}{s \spr (1-d)}\left( \frac{1}{2}(s + \spr)  \Gamma^{\al}_{\al}(s,\spr) - d\, \Gamma(s,\spr) )   \right)  \;,
\end{equation}
and the  analogous expression of \eqref{eq:rhoPbq} is lengthy for the Kaon and is given in a Mathematica ancillary notebook attached to the arXiv version. 

Changing the prescription \eqref{eq:pres} by 
 $\frac{1}{2}(s + \spr) \to \sqrt{s \spr}$ results in a $15\%$-change which is sizeable but 
not extremely large and well within the error. 
In addition we use a weight function $1/s\spr$ as described in \APP\ref{app:winv} as otherwise the daughter sum rule is off by at least a factor of two which is very large in view of how well it works 
in all other cases. 
 
Proceeding as before we obtain 
the following values  
\begin{equation}
\label{eq:SRaxial}
\{  s_0, \hat{M}^2 \}_{K}   = \{  0.7(1), 0.95(0.5)  \}  \GeV^2     \;,  \quad 
 U(s_0 \pm 0.1,M^2)_{\Delta m_K|_{\text{QED}} } = 1.00 \pm 0.10 \;,
\end{equation} 
for the sum rule parameters  and the daughter sum rule \eqref{eq:U}.
Using the input parameters in \TAB\ref{tab:inputParams}, the $f_{K}$ sum rule to LO (cf. \APP\ref{app:decayConst})
 and 
\eqref{eq:SRaxial} we get
\begin{equation}
\Delta m_K|_{\text{QED}} =  +1.85^{+0.42}_{-0.66} \MeV \;.
\end{equation}
Scale dependent quantities are evaluated at $\mu = 2\GeV$. The uncertainty again comes from adding individual errors in quadrature. The dominant uncertainty ($75\%$) comes from the $m_s$ mass with the remaining uncertainty due to the  the duality parameter $a$ in \eqref{eq:de}.

As stated in the introduction, the pion proved more difficult. That is we were not able to find stable sum rules
satisfying the daughter sum rule for reasonable values of the 
continuum threshold.\footnote{The extra disconnected diagram for the $\pi^0$, e.g. \cite{Feng:2021zek}, is small since the $\ga_5$ generates a Levi-Civita tensor which enforces two extra loops. This is reflected in the smallness of the lattice result  \cite{Feng:2021zek} and also by the fact that the LO chiral Lagrangian does not contribute to $\pi^0$ 
(cf.\,\APP\ref{app:DeltaPi}).}
We believe that is due to its small mass $m_\pi$ which is considerably below the other hadronic masses. Conversely the Kaon mass, while being a pseudo-Goldstone, is much closer to the other hadrons 
(due to $m_s$ being close to $\Lambda_{\text{QCD}}$). 

\section{Linear Quark Mass Correction $\Delta m_H|_{m_q}$}
\label{sec:mq}

As stated in the introduction (and cf.\,\APP\ref{app:FH}) the $\ORD(m_q)$-corrections 
are governed by $ \matel{{H}}{\bar qq}{{H}} $ \eqref{eq:DelmBmq}.
 For the $B,D$-meson we compute this matrix element 
from QCD sum rules in \SEC\ref{sec:DelmqHEAVY}, using similar techniques as for the QED correction, and  
for  light mesons we resort to soft theorems cf. \SEC\ref{sec:DelmqLIGHT} as the corresponding sum rules are inferior.

\subsection{QCD Sum Rule Computation of $\matel{\bar H}{\bar q q}{\bar H}$ for $H = B,D$}
\label{sec:DelmqHEAVY}

\begin{figure}[ht]
 \centerline{
\begin{overpic}[width=5.0in]{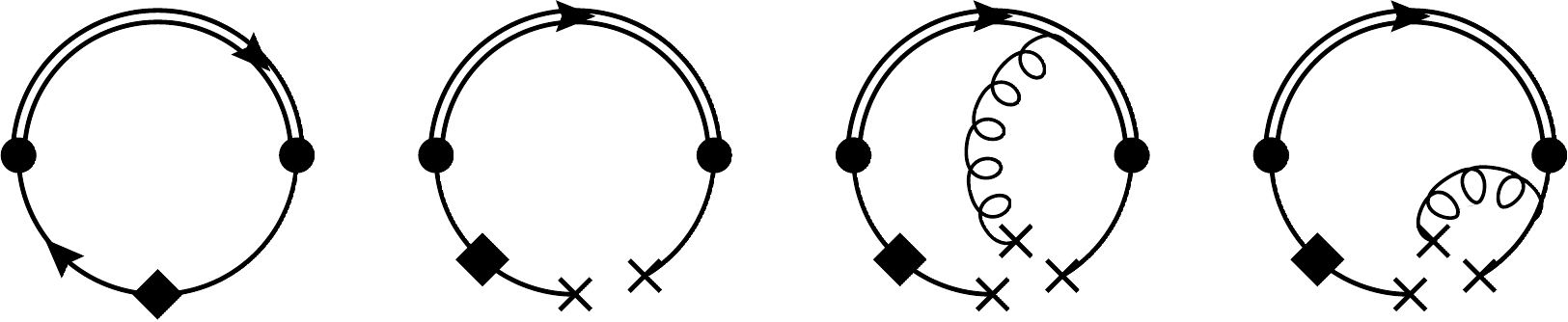} 
\put(5,14){$b$}
\put(5,5.5){$\bar q$}
\put(65,10){$g$}
\end{overpic}}
 \caption{\small  Diagrams contributing to  the matrix element 
 $ \matel{{B}}{\bar qq}{{B}} $.  They are analogous to the ones 
 in \FIG\ref{fig:diaQED} but the square blob denotes the insertion of the $\bar qq$-operator. 
 Perturbation theory is minimal and the quark condensate diagram is the main contribution. 
 The mixed condensate diagrams $\VEV{\bar q G q}$ are mainly useful to stabilise the sum rule. 
 }
 \label{fig:diaQCD}
 \end{figure}

 In order to anticipate the hierarchy of  diagrams shown in \FIG\ref{fig:diaQCD} it is worthwhile to contemplate on the heavy quark behaviour. The matrix element scales like $(H = B)$  for definiteness).  
\begin{equation}
\label{eq:rel}
\matel{B}{\bar qq}{B}  = \ORD(m_b) \;,
\end{equation}
 for relativistically  normalised states, 
$\langle B(p) | B(q) \rangle  = 2 E_B(\vec{p}) (2\pi)^3  \de^{(3)}(\vec{p} -  \vec{q})$,  due to the factor $E_B = \ORD(m_b)$.  On the one hand, the operator $\bar qq$  demands a chirality flip in perturbation theory 
and this cannot come from the $m_b$-mass since the latter is entirely kinematic as we have just established. 
On the other hand the condensate contribution itself $\VEV{\bar qq}$ does not require this flip and is 
therefore unsuppressed and numerically leading.

To do the computation we start from the following  correlation function 
\begin{equation}
\Pi(p^2, \tilde{p}^2,r) = i^2 \int_{y,z} e^{i (\tilde{p} z-p y-xr)}  \matel{0}{T J_B^\dagger (z) (\bar q q)(x) J_B(y)}{0} \;,
\end{equation}
where $J_B$ has been defined in \eqref{eq:JB} and the auxiliary momentum $r$ takes on the same r\^ole 
as before. 
The  double dispersion relation  of  the correlation functions reads 
\begin{equation}
\label{eq:QCDqqdispersion}
 \Pi(p^2, \tilde{p}^2,r) =  \int   \frac{ds d\spr \, \rho_{\Pi}(s,\spr)  }{( s- p^2 -i0) 
( \spr- \tilde{p}^2 -i0) }  
=  \frac{Z_B^2 \matel{\bar B}{\bar q q}{\bar B}}{( m_B^2- p^2) 
( m_B^2 - \tilde{p}^2) } + \dots \;.
\end{equation}
with $( 2\pi i)^2  \rho_{\Pi}(s,\spr) =  \text{disc}_{s,\spr} [\Pi(s,\spr) ] $, and the matrix element is then 
given by
\begin{equation} 
 \matel{\bar B}{\bar q q}{\bar B} = \frac{1}{Z_B^2}  \int_{m_+^2}^{\bar{\de}^{(a)}(m_+^2)} ds\, e^{\frac{(m_B^2-s)}{M^2}}   \int_{m_+^2}^{\bar{\de}^{(a)}(s)}  d\spr  \, e^{\frac{(m_B^2-\spr)}{M^2}} 
 \rho_{\Pi}(s,\spr)  \;,
\end{equation}
with $\bar{\de}^{(a)}$ defined in \eqref{eq:de}. 
The three contributions depicted in \FIG\ref{fig:diaQCD} are described below.
\begin{itemize}
\item \emph{Perturbation theory} is given by
\begin{equation}
\rho_{\Pi}(s,\spr) =  \frac{ \mplus^2 N_c m_q}{2 \pi^2} \, \frac{s-(m_b - m_q)^2}{s+m_q^2 - m_b^2}   \lambda^{\frac{1}{2}} \delta(\spr - s) \;,
\end{equation}
with the anticipated $\ORD(m_q)$-suppression.  This term is negligible.
\item \emph{The $\langle \bar q q \rangle$ condensate} evaluates to 
\begin{equation}
\matel{\bar B}{\bar q q}{\bar B} = - \frac{ 4 \mplus^2 m_b^2 \langle \bar q q \rangle}{Z_B^2} e^{\frac{2(m_B^2 - m_b^2)}{M^2}} \;,
\end{equation}
which is not suppressed by $\ORD(m_q)$ and thus dominant.
\item \emph{The mixed condensate} yields
\begin{alignat}{2}
& \matel{\bar B}{\bar q q}{\bar B} &\;=\;&  - \frac{ \mplus^2 \langle \bar q \sigma s_g g G q \rangle}{Z_B^2}  e^{\frac{2(m_B^2 - m_b^2)}{M^2}}\left(  ( 1- \frac{3m_b^2}{M^2}) + ( \frac{5}{8} + \frac{2 m_b^2}{M^2} - 
 \frac{4 m_b^4}{M^4} )   \right) \;,
 \end{alignat}
 which is not suppressed either as it is in the same chirality representation as the quark condensate.  
 The first and second term in round brackets are from the third and fourth diagram 
 in \FIG\ref{fig:diaQCD}. 
\end{itemize}
We consider it worthwhile to comment how the lack of $m_q$-suppression in the condensate 
contribution arises. 
Its origin is the   propagator $1/(r^2 - m_q^2+i \eps)$ (we work in the $\vec{r} =0$ frame) 
\begin{equation}
r^2 - m_q^2 + i \eps = ( \sqrt{s} - (\sqrt{\spr}+m_q - i \eps'))( \sqrt{s} - (\sqrt{\spr}-m_q+i \eps')) \;,
\end{equation}
 which when cut gives a term of the form $\frac{\sqrt{s}}{m_q} \delta(s-(\sqrt{\spr}+m_q)^2)$. The $1/m_q$ thus removes the $\ORD(m_q)$-suppression in the numerator. 
Numerically perturbation is entirely negligible and this is also the reason for not including 
the gluon condensate which is expected to be further suppressed $\ORD(\Lambda_{\text{QCD}}^4/M^4)$ as compared to perturbation theory. 

\subsubsection{Numerics}
The basic procedure for the numerics is the same as described in \SEC\ref{sec:numerics}. However, the choice of scheme is not as important in this case. Any of the schemes, pole, kinetic and $\MSbar$ give similar results and indicate stability.  
The situation is certainly clearer with respect to the $m_b$-mass itself as the matrix element is $\ORD(m_b)$ \eqref{eq:rel} 
and $\Delta m_B|_{m_q}$ itself  is $\ORD(m_b^0)$ whereas $\Delta m_B|_{\text{QED}}$ is computed from a non-local correlation function where the 
$m_b$-dependence is more difficult to track.
Since the perturbative contribution is suppressed, there is no $s_0$ dependence (there would be at NLO in $\al_s$). Hence we can fix the Borel value $M^2$ to satisfy the daughter sum rule \eqref{eq:U}, obtaining the following sum rule parameters 
\begin{equation}
\label{eq:SRmq}
\{  s_0, \hat{M}^2 \}_{B}   = \{  35.0, 4.0\}  \GeV^2  \;, \quad \{  s_0, \hat{M}^2 \}_{D}   = \{  6.0, 0.75 \}  \GeV^2   \;,
\end{equation}
and daughter sum rules
 \begin{alignat}{2}
 \label{eq:Uqq}
&  U(s_0,\hat{M}^2 \pm 0.15 \GeV)_{\Delta m_B|_{m_q} } &\;=\;& 1.00^{+0.03}_{-0.02}   \;,    \nonumber \\[0.1cm]
& {U}(s_0,\hat{M}^2 \pm 0.05 \GeV)_{\Delta m_D|_{m_q} } &\;=\;& 1.00^{+0.20}_{-0.12} \;.
 \end{alignat}
Using the input parameters in \TAB\ref{tab:inputParams} (with $m_b^{kin}(1\GeV),\bar{m}_c(\bar{m}_c)$), 
the $f_{B,D}$ sum rule to LO (cf. \APP\ref{app:decayConst}) and \eqref{eq:SRmq} we get 
\begin{equation}
\label{eq:BqqB}
\matel{\bar B}{\bar q q}{\bar B}_{\mu = 1 {\tiny \GeV}} = 5.99^{+1.99}_{-1.41} \GeV \;, \quad \matel{\bar D}{\bar q q}{\bar D}_{\mu = \bar{m}_c {\tiny \GeV}} = 3.40^{+1.78}_{-1.71} \GeV \;,
\end{equation}
for the matrix elements and 
\begin{equation}
\label{eq:RZ}
\Delta m_B|_{m_q} = -1.88^{+0.49}_{-0.71} \MeV \;, \quad  \Delta m_D|_{m_q} = +2.68^{+1.48}_{-1.38} \MeV \;,
\end{equation}
for the mass differences. 

As this is a LO computation the errors are large, primarily coming from $M^2$ with a small contribution ($20\%$) from the light quark masses. Note that the set value of $M^2$ is not independent of higher order $\alpha_s$ corrections. For the $D$-meson especially, the convergence of the sum rule is not good.
This is reflected in the mixed condensate  contributing a  sizeable $20\%$-uncertainty.

\subsection{$SU(3)_F$ estimates of $\matel{\bar H}{\bar q q}{\bar H}$ for $H = B,D$}
\label{sec:DelmqSU(3)}

Alternatively, one may use $SU(3)_F$ flavour symmetry $\matel{B}{\bar qq}{B}  \approx \matel{B_s}{\bar ss}{B_s}$ 
to estimate  $\matel{B}{\bar qq}{B}$ \cite{Colangelo:1997tc}. Following this analysis one may write 
($m_{ud} \equiv \frac{1}{2}(m_u + m_d)$)
\begin{equation}
(2 m^2_{B_s} - m^2_{{B^+}}-m^2_{{B^0}}  )  = 2 (m_s -m_{ud})  \matel{B}{\bar qq}{B}  \;,
\end{equation}
from which
\begin{equation}
  \matel{B}{\bar qq}{B}  \approx \frac{m^2_{B_s} - m_B^2}{(m_s -m_{ud})} \;,
\end{equation}
follows. Employing the input from the PDG \cite{PDG}  this leads to\footnote{Or taking the $\eta \to 3 \pi$ analysis \cite{Colangelo:2018jxw}, which in this case makes a difference,  results in
\begin{equation}
\Delta m_B|_{m_q} = -2.54^{+0.17}_{-0.18}  \pm 20\%_{SU_3}\MeV \;, \quad  \Delta m_D|_{m_q} = +3.01^{+0.21}_{-0.20}  \pm 20\%_{SU_3} \MeV \;, 
\end{equation}
a more precise result.}  
\begin{equation}
\label{eq:Italia}
\Delta m_B|_{m_q} = -2.37^{+0.35}_{-0.43} \pm 20\%_{SU_3} \MeV \;, \quad  \Delta m_D|_{m_q} = +2.81^{+0.51}_{-0.41} \pm 20\%_{SU_3} \MeV \;.
\end{equation}
 We have added a characteristic $20\%$ $SU(3)_F$-violation due to the use 
of the $\matel{B}{\bar qq}{B}  \approx \matel{B_s}{\bar ss}{B_s}$. 
The result are well compatible with  
\eqref{eq:RZ} and we shall not use them any further.  Note that in the 
heavy quark limit we have $\Delta m_B|_{m_q} = -  \Delta m_D|_{m_q}$ since the $c$ and $b$ are up and down quark types respectively.  This heavy quark limit relation  holds  reasonably
as already  observed in \cite{Colangelo:1997tc} (with slightly different input).

\subsection{Soft Goldstone estimate of $\matel{ L}{ \bar q q}{ L}$ for $L = \pi,K$}
\label{sec:DelmqLIGHT}

The matrix elements  $\matel{ L}{\bar q q}{ L}$ where $L = \pi, K$ is a pseudo-Goldstone boson
may be estimated using soft-pion techniques which in this case lead to the famous 
GMOR-relation \cite{Gell-Mann:1968hlm}.  Concretely \cite{Donoghue:1992dd}
\begin{equation}
\label{eq:SU(3)}
 m_{\pi^{+,0}}^2 =    (m_u + m_d) B_0 \;, \quad 
 m_{K^+}^2  =    (m_u+ m_s) B_0  \;, \quad m_{K^0}^2  =    (m_d+ m_s) B_0 \;,
\end{equation}
which are to first order in the quark masses, with no QED corrections and the constant is
$B_0 = - \frac{ 2 \VEV{  \bar qq}}{f_\pi^2} \approx 2.26 \GeV$ at $\mu = 2\GeV$. We see that  for the pions there is no difference 
to linear order which is a consequence of isospin \cite{Donoghue:1996zn}. The pion mass 
splitting is a $\Delta I = 2$ isospin effect since the relevant matrix element has two pion states where
the quark masses themselves are of  $\Delta I =1$.  Hence it takes at least two powers of the quark mass difference. Fortunately, the latter follows in a straightforward manner from chiral perturbation theory and one obtains to LO
\begin{alignat}{3}
\label{eq:PiKaon}
&  \Delta m_K|_{m_q}   &\;=\;&   \frac{  m_u - m_d}{m_s - m_{ud}} \frac{m_K^2 - m_\pi^2}{2m_K}  =  \frac{  m_u - m_d}{ 2 m_{ud}}  \frac{m_\pi^2}{2m_K}    
&\;=\;& -6.74^{+0.98}_{-1.21}   \MeV  \;,    \nonumber \\[0.1cm]
&  \Delta m_\pi|_{m_q}   &\;=\;& \frac{1}{16}  \frac{  m_d - m_u}{m_s - m_{ud}}  \frac{  m_d - m_u}{ m_{ud}}   {m_{\pi  }}  &\;=\;& + 0.16^{+0.06}_{-0.05} \MeV   \;,
\end{alignat}
using the values from the PDG \cite{PDG}.  As expected the pion contribution is rather small as a result of being second order in the quark mass difference. It is noteworthy that one obtains $ \Delta m_K|_{m_q} \approx -5.7 \MeV$ when using \eqref{eq:SU(3)} directly which can be seen as a $SU(3)_F$ correction 
 which is well covered by the quoted uncertainty.
 
\section{Final Overview and Conclusions}
\label{sec:conclusions}

In this paper we have computed the mass difference of the charged and neutral  $B$-, $D$- and 
$K$-mesons.  The results, which originate from electromagnetic and quark mass effects, are summarised 
and contrasted with experimental values in \TAB\ref{tab:overview}. 
The electromagnetic contribution is computed from the second order formula \eqref{eq:DelmBQED} 
in \SEC\ref{sec:QED} and may be regarded as the core part of this paper.  
$ \Delta m_\pi|_{\text{QED}}$ is taken from a soft-pion theorem (cf.\,\APP\ref{app:DeltaPi})
for completeness and comparison.
Quark mass effects are obtained  from the Feynman-Hellman formula \eqref{eq:DelmBmq} and its corresponding matrix element is computed in \SEC\ref{sec:DelmqHEAVY} for the $B$ and the $D$ respectively 
whereas for the $K$ and the $\pi$  a soft theorem  turns out to be more reliable. 

The results obtained are consistent with the current experimental values.
The uncertainties are above $20\%$ and indeed more cannot be expected from a double dispersion sum rule  at leading order in the strong coupling constant. Experimental uncertainties 
are one or two orders of magnitude lower. 
 
 \begin{table}[h]
\renewcommand*{\arraystretch}{1.2}
\begin{center}
\begin{tabular}[h]{l |  r r  r | r }
  $H$  &   $ \Delta m_H|_{\text{QED}}$   &  $ \Delta m_H|_{m_q}$  & $ \Delta m_H$   & 
  $ \Delta m_H|_{\text{PDG} \mbox{\cite{PDG}}}$
    \\  \hline 
   $B$ & $+1.58(24)\,\MeV$ & $-1.88(60)\,\MeV^{\,a}$ & $-0.30(65)\,\MeV$  &$ -0.32(5)\,\MeV$ \\
    $D$ & $+2.25(70)\,\MeV$  & $+2.7(1.4)\,\MeV^{\,a}$  & $+4.9(1.6)\,\MeV$ & $+4.822 (15)\,\MeV$  \\
     $K$ & $+1.85(54)\,\MeV$ &  ${\it -6.7(1.1)\,\MeV^{\,b}}$ &   $-4.9(1.2) \,\MeV$ & $-3.934(20)\,\MeV$  \\
     $\pi$ & \  ${\it +4.8(1.2)\,\MeV^{\,c} } $  &  ${\it +0.16(5)\,\MeV^{\,b}}$ &   $+5.0(1.2)\,\MeV$  & $+4.5936(5)\,\MeV$ \\
\end{tabular}
\end{center}
\caption{\small Our values of $\Delta m_H$ due to the electromagnetic mass difference 
and the quark masses compared  to the PDG values. 
The entries marked with $^{a}$ are obtained from the $\matel{H}{\bar qq}{H}$ matrix element 
in conjunction with the  Feynman-Hellman theorem 
(valid to LO in $m_{q}$). The values in italic should \emph{not} be regarded as predictions of this work.
E.g. $^{b}$derived from the soft theorem for (pseudo-) Goldstone bosons (cf.\,\APP\ref{sec:DelmqLIGHT}) and
$^{c}$results from soft theorem 
in conjunction with the  Weinberg sum rules (cf.\,\APP\ref{app:DeltaPi}).
It is noteworthy that   
$\Delta m_\pi|_{m_q} = \ORD( (m_u- m_d)^2)$ which explains its smallness.  
For comparison some lattice values $\Delta m_D = 5.47(53) \MeV$  and $\Delta m_K = -4.07(15)(15) \MeV$ \cite{Giusti:2017dmp} and $\Delta m_D = 4.68(10)(13) \MeV$ \cite{Borsanyi:2014jba} which are of course more precise as the lattice is suited for mass determination, even in the presence of QED, 
and due to the full inclusion of QCD.}
\label{tab:overview}
\end{table}
 
 The values in \TAB\ref{tab:overview} deserves some comments as they are not easily guessed by rules of
  thumb by a practitioner in non-perturbative QCD.  The parametric estimate of
 $\Delta m_H|_{\text{QED}} = c \, Q^\text{eff}_{H} \frac{\al}{\pi} \Lambda_{\text{QCD} }$ 
 with $ \Lambda_{\text{QCD}} = 200\MeV$ and $Q^\text{eff}_{D} = 2Q^\text{eff}_{B,K} = 2/3$, leads to $c \approx 10$-$20$
 which is a rather large number. 
 To put this into perspective, 
 one should keep in mind that these kind of estimates are not 
 straightforward as the mass difference is obtained from a non-local (long distance) correlation function 
 \eqref{eq:DelmBQED}.
 The scale for the quark mass effect is of course set me $m_u - m_d \approx 2.5\MeV$ and its sign depends 
 on whether the non $q=u,d$ quark is of the up (charm) or down (beauty, strange) type quark. 
 The cancellation to almost an order of magnitude of the electric and the quark mass contribution for the $B$-meson is remarkable, leading to an inflated uncertainty in  
$\Delta m_B$. 

The main aim of this paper was to show that it is possible to understand the isospin 
mass difference from QCD sum rules, that is to obtain values  compatible with experiment.    
The sum rule computation could be improved by including radiative corrections in the strong coupling 
constant which would be a formidable  task.   
 Perhaps more interestingly,   the formalism developed in this paper could be applied 
to  baryons to obtain  the proton-neutron mass difference for instance.

\acknowledgments

RZ is supported by a CERN associateship and
an STFC Consolidated Grant, ST/P0000630/1.   
We are grateful to Michele Della Morte, Antonin Portelli and Max Hanson for informative comments on the lattice literature. 

\appendix

\section{Variants of Quark-Hadron Duality}
\label{app:duality}

In this appendix we elaborate on variations of quark-hadron duality. 
This is best explained by example. Consider the axial correlator in connection with the $K$
\begin{equation}
\label{eq:AX}
\Pi_{\al \be} = i \int d^4 x e^{i px} \matel{0}{T A^\dagger_\al(x) A_\be(0)  }{0}  = p_{\al} p_{\be} \Pi(p^2) + g_{\al\be} \hat{\Pi}(p^2) \;,
\end{equation}
with $A_\be$ defined in \eqref{eq:A}. The Kaon appears in the first structure 
\begin{equation}
\label{eq:basic}
 \Pi(p^2) = \frac{f_K^2}{m_K^2 - p^2}  + \dots \;,
\end{equation} 
where the dots stand for higher states as usual. 
QCD sum rules consists of two steps. Firstly the observation that 
\begin{equation}
\label{eq:eq}
\Pi(p^2) \approx \Pi(p^2)_{\textrm{pQCD}} \;,
\end{equation}
for some $p^2$ outside the physical region (could be $p^2 < 0$), 
where pQCD stands for perturbative QCD with OPE improvements.  
In  a second step one rewrites \EQ\eqref{eq:eq} as a dispersion relation 
followed by a Borel transform under which $(s- p^2)^{-1} \to \exp(-s/M^2)$ ($M^2$ is the Borel parameter) 
which results in
\begin{equation}
\label{eq:semiglobal}
\int_0^{\infty} e^{-s/M^2} \rho(s) \approx \int_0^{\infty} e^{-s/M^2} \rho_{\textrm{pQCD}}(s) \;,
\end{equation}
with $\rho(s) = \frac{1}{2 \pi i} \textrm{disc}_s \Pi(s) = f_K^2 \de(s-m_K^2) + \dots$ and the pQCD part is defined analogously. 
The one assumption is then that this integral can be broken up as follows 
\begin{equation}
\label{eq:semiglobal}
\int_0^{s_0} e^{-s/M^2} \rho(s) \approx \int_0^{s_0} e^{-s/M^2} \rho_{\textrm{pQCD}}(s) \;,
\end{equation}
and \eqref{eq:semiglobal} is sometimes referred 
to as semi-global quark hadron duality \cite{Shifman:2000jv}.
One way to determine $s_0$ is to impose the daughter sum rule \eqref{eq:DSR}
and then for consistency with the  duality assumption $s_0$ ought to be somewhere between $(m_K + 2 m_\pi)^2$ and $(m_K + 4 m_\pi)^2$.

 We want to briefly contemplate for which types of weight functions $\omega(s)$
  \eqref{eq:semiglobal}  
\begin{equation}
\label{eq:semiglobal2}
\int_0^{s_0} e^{-s/M^2} \rho(s)  \omega(s)  \approx \int_0^{s_0} e^{-s/M^2} \rho_{\textrm{pQCD}}(s)  \omega(s) \;,
\end{equation}
with  corresponding  \eqref{eq:DSR}
 \begin{eqnarray}
\label{eq:DSR2}
m_B^2 =    \int_{\textrm{cut}}^{s_0} e^{-s/M^2}  \rho_{\textrm{pQCD}}(s) \omega(s) \,s\, ds  / ( \int_{\textrm{cut}}^{s_0}  e^{-s/M^2}  \rho_{\textrm{pQCD}}(s)   \omega(s)  ds)  \;,
\end{eqnarray}
can hold. The crucial point is to be able to justify the analogue of \EQ\eqref{eq:eq}.

\subsection{Weight function $\omega (s) =s$}

We might start by rewriting the $p_\al p_\be$-part in \eqref{eq:AX} as follows
\begin{equation}
p_{\al} p_{\be} \Pi(p^2)=   \frac{p_\al p_\be}{p^2} (p^2 \Pi(p^2) ) \;.
\end{equation}
For the pQCD part one may directly write $ \rho_{\textrm{pQCD}}(s) \to s \rho_{\textrm{pQCD}}(s)$ since $p^2$ does not lead to new singularities. 
Using \eqref{eq:basic}, the  QCD part can be written as
\begin{equation}
(p^2 \Pi(p^2) ) = p^2 \frac{f_K^2}{m_K^2-p^2} + \dots = - f_K^2 + m_K^2 \frac{f_K^2}{m_K^2-p^2} + \dots \;,
\end{equation}
where $-f_K^2$ is a constant that will disappear under Borel transformation and thus $ \rho(s) \to s \rho(s)$ works the very same way. The analogue of $\eqref{eq:eq}$ can be justified in this case by 
replacing $ A^\dagger_\al(x) \to -  \partial^2A^\dagger_\al(x)$ \eqref{eq:AX}.\footnote{In our case this is not trivial as $A^\dagger_\al$ is not QED gauge invariant but it can still be used at LO. 
In the general case this requires more thought.} 
Weight functions of polynomials are generally referred to as moments and
are familiar to the community  e.g. moments 
in $b \to c \ell \nu$ for example \cite{Bigi:1997fj}. It is quite clear that one can not take arbitrarily high powers of moments as then duality will be challenged since smoothness is lost.

\subsection{Weight function $\omega(s) = \frac{1}{s-\eta}$}
\label{app:winv}
 
 Choosing a weight function 
 \begin{equation}
 \label{eq:Iweight}
  \omega(s) = \frac{1}{s-\eta} \;,
 \end{equation}
 is equivalent to working with 
 a subtracted dispersion relation fo the form 
\begin{equation}
\label{eq:hm}
 \frac{\Pi(p^2)-\Pi(\eta)}{ p^2-\eta} =  \int \frac{ds \rho(s)}{(s-p^2)(s-\eta)} + c \;,
 \end{equation} 
where  $c =  - \int {ds \rho^A(s)/(s(s-\eta))}  +\Pi^{'}(\eta) $ is a subtraction constant such that the limit $p^2 \to 0$ comes out correctly. The constant $c$ is though not important in the end as it vanishes under Borel transformation.  The question of whether one can use \eqref{eq:Iweight} then turns into the question whether the left hand side can be computed reliably. 

In our application to Kaons we have chosen $\eta  = 0$ which is close but still below the Kaon resonance. 
We have checked that for the $f_K$ sum rule with  $s_0 = 0.7 \GeV^2$ 
 the agreement is reasonable and this serves at least as a partial justification of the procedure 
 in \SEC\ref{sec:KQED}.

\section{Numerical Input}
\label{app:input}

The numerical QCD input  is summarised in \TAB\ref{tab:inputParams} and below 
we give the numerical values of the the decay constant from sum rule which are the effective 
LSZ factors.

\begin{table}[btp]
\addtolength{\arraycolsep}{3pt}
\renewcommand{\MeV}{\,{\textrm{MeV}}}
\renewcommand{\GeV}{\,{\textrm{GeV}}}
\renewcommand{\arraystretch}{1.3}
\resizebox{\columnwidth}{!}{
\begin{tabular}{c|c|c|c|c|c} 

\multicolumn{6}{c}{$J^P =0^- \mbox{ Meson masses~\cite{PDG}}$}\\\hline
$\mB$            & $\mBs$         & $m_D$       & $\mDs$       & $m_K$ & $m_\pi$  \\\hline
$5.280  \GeV$ & $ 5.367\GeV$ & $1.867 \GeV $ & $1.968 \GeV$ & $0.496 \GeV$ & $0.137 \GeV$ \\\hline

\multicolumn{6}{c}{$J^P =0^- \mbox{ Mass Differences~\cite{PDG}}$}\\\hline
$\Delta m_B$            & $\Delta m_D$         & $\Delta m_K$       & $\Delta m_\pi$       & $$ & $$  \\\hline
$-0.32(5)  \MeV$ & $ +4.822 (15)\MeV$ & $-3.934(20) \MeV $ & $+4.5936(5) \MeV$ & $$ & $$ \\\hline

 \multicolumn{6}{c}{$\mbox{Quark masses~\cite{PDG}}$}\\\hline
$\bar m_b(m_b) $    & $\bar m_c(m_c) $      & $m_b^{\textrm{pole}}$ & $m_c^{\textrm{pole}}$ & 
$m_b^{kin}{\lscale{1}}$ & $m_c^{kin}{\lscale{1}}$ \\\hline
$4.18^{+0.03}_{-0.02} \GeV $&$ 1.27(2) \GeV $  &$ 4.78(6)\GeV   $     & $1.67(7) \GeV $       & $4.53(6)\GeV $  & 1.13(5)\\\hline
$\bar m_{s\lscale{2}}$  & $\bar m_{d\lscale{2}}$    & $\bar m_{u\lscale{2}} $      & $\bar{m}_{ud}{\lscale{2}}$ & 
$\frac{\bar{m}_u}{\bar{m}_{d}}$ & $\frac{\bar{m}_s}{\bar{m}_{ud}}$ \\\hline
$93.4^{+8.6}_{-3.4}  \MeV$   & $4.67^{+0.48}_{-0.17} \MeV $ & $2.16^{+0.49}_{-0.26} \MeV $  & $3.45^{+0.35}_{-0.15} \MeV  $  & $0.474^{+0.056}_{-0.074} $ & $27.33^{+0.67}_{-0.77}$ \\\hline

 \multicolumn{6}{c}{$\mbox{Condensates}$}\\\hline

$\qbarq_\lscale{2} \mbox{~\cite{Bali:2012jv}} $ & $\sbars_\lscale{2} \mbox{~\cite{McNeile:2012xh}} $ &  $m_0^2 \mbox{~\cite{Ioffe:2002ee}}$ & $  \vev{\frac{\alpha}{\pi}G^2}\mbox{~\cite{SVZ79I}} $
  & $$ & $$  \\\hline
$-(269(2) \MeV)^3           $                  & 1.08(16) $\qbarq$                                             & $0.8(2)\GeV^2    $ 
& $0.012(4)\GeV^4  $  
 & $$ & $$               \\\hline
\end{tabular}
}
\caption{\small Summary of input parameters. Note as inputs into the sum rules we use $m_H = m_{H^-} $, as which has a completely negligible impact. The quantity $m_{ud} \equiv \frac{1}{2}(m_u+m_d)$ is the light quark average. The mixed condensate is parameterised 
as $\langle \bar q \sigma s_g g G q \rangle = m_0^2 \VEV{\bar qq}$ as is standard in the literature.}
\label{tab:inputParams}
\end{table}

\subsection{Decay constants $f_B$, $f_D$ and $f_K$}
\label{app:decayConst}

The extraction of both the QED mass shifts and the linear quark mass corrections, require values for the decay constants $f_B$, $f_D$ and $f_K$. Note that, for consistency with the rest of this paper
 these are evaluated  at LO in QCD. The LO expressions for the pseudoscalar ($B,D$) and axial ($K$) correlators are well known (e.g. \cite{Jamin:2001fw,Ball:2005vx}). The following values
\begin{alignat}{6}
f_B &=& \enspace & 0.157\GeV \, , \qquad \{s_0,M^2\} & \enspace=\enspace& \{33.5,6.0\} &\GeV^2  \;,\nonumber \\[0.1cm] 
f_D &=& \enspace & 0.158\GeV \, , \qquad \{s_0,M^2\} &\enspace=\enspace& \{5.7,2.0\} &\GeV^2\;, \nonumber \\[0.1cm] 
f_K &=& \enspace  & 0.147\GeV \, , \qquad \{s_0,M^2\} &\enspace=\enspace& \{1.1,1.5\} &\GeV^2 \;,
\end{alignat}
are obtained. 

\section{Self Energies and Condensates for $\Delta m_H|_{\text{QED}}$}
\label{app:self}

In this appendix we present some extra computations: the self energies and condensate contributions to $\Delta m_B |_{\text{\text{QED}}} $. These are important for stabilising the sum rules 
but do not affect the actual value of  $\Delta m_B |_{\text{\text{QED}}} $ per se. This is the case since graphs proportional to $Q_b^2$ are cancelled in the mass difference. The only non-zero graph contributing to the mass shift is the $q$-$q$ self energy, but it is numerically negligible. 
We wish to note that in all these graphs explicit gauge independence has been verified to hold after the double-cut is taken.

\subsection{Perturbation theory} 

The perturbative $b$-$b$ self energy graph, after mass renormalisation, takes on the form
\begin{equation}
\rho_{\Gamma_{bb}}(s,\spr) = \frac{N_c m_+^2 Q_b^2 \alpha }{32\pi^3 m_B} \cdot \lambda^{\frac{1}{2}} \cdot \frac{s - m_-^2}{s + m_+ m_-} f^{\text{R}}(m_b^2)  \delta(\spr - s)  \;,
\end{equation}
with the renormalised $f^{\text{R}}$\footnote{Note that the vanishing in the pole scheme is clear, by the very definition of the scheme, since we are on-shell after the cuts.} 
\begin{equation}
f^{\text{R}}(m^2) = f(m^2) + \frac{32\pi^2 m^2}{e^2} \de Z_{m} =  
\begin{cases}
2m^2 \bigg( 4 + 3 \ln \frac{\mu^2}{m^2}  \bigg),  & \overline{MS} \\
0, & \text{Pole} \\
2m^2 \bigg( \frac{16 \mu}{3m}+ \frac{2 \mu^2}{m^2} \bigg), & \text{Kinetic}
\end{cases}
\end{equation}
\begin{equation}
f(m^2) = 4m^2 B_0(m^2,0,m^2) + (d-2)A_0(m^2) \;.
\end{equation}
The functions $A_0$ and $B_0$ are the standard Passarino-Veltman functions with (\textsc{FeynCalc}) normalisation $(2\pi \mu)^{2 \epsilon} \int \dd[d]{k} / (i \pi^2)$. Explicitly these are
\begin{equation}
B_0(m^2,0,m^2) =  \frac{1}{\hat{\epsilon}} + 2
 + \log ( \frac{\mu^2}{m^2} ) ,  \qquad A_0(m^2) = m^2 \bigg( \frac{1}{\hat{\epsilon}} + 1
 + \log ( \frac{\mu^2}{m^2} ) \bigg) \;,
\end{equation}
with $\frac{1}{\hat{\epsilon}} = \frac{1}{\epsilon} - \gamma_E + \log 4\pi$. The $q$-$q$ graph can be obtained by replacing $b \rightarrow q$ in the result and since it is $\ORD(m_q^2)$ it is negligible. 

\subsection{Condensates}
\label{app:cond}

The only relevant condensate graph is given in \FIG\ref{fig:diaQED} ($4^{\text{th}}$ diagram). With $m_q \rightarrow 0$ the density is
\begin{equation}
\rho^{\qbarq}_{\Gamma_{bb}} = - \frac{m_b^2 \alpha Q_b^2}{8 \pi m_B} m_b \qbarq  \delta(s-m_b^2) \delta(\spr - m_b^2) f^{\text{R}}(m_b^2) \;.
\end{equation}
Light quark mass corrections come from Taylor expanding the quark fields, leading to derivatives of  
$\de$-functions. It is thus more convenient to directly display the resulting mass shift
\begin{equation}
 \Delta m_B|_{\qbarq}   =  - \frac{m_+^2 \alpha Q_b^2}{8 \pi m_B Z_B^2}  e^{\frac{2(m_B^2 - m_b^2)}{M^2}} \qbarq  \bigg( m_b - \frac{m_q}{4} \big(1+\frac{4m_b^2}{M^2} \big) \bigg)  f^{\text{R}}(m_b^2) 
\end{equation}
The $\qbarq$ condensate graph where the photon connects the $b$ and the $q$-quark is not of short distance type (it leads to $1/m_q^2$ in the propagator) and is therefore omitted. This is similar to the $B \to \ga$ form factor although in 
that case the physics is covered by the photon distribution amplitude (e.g. \cite{Janowski:2021yvz}).

\section{Some Classic Results}
\label{app:classic}

In this appendix we summarise some classic results which are of use and referred to in the paper.

\subsection{Linear quark mass dependence from Feynman-Hellman theorem}
\label{app:FH}

In order to derive the  Feynman-Hellman theorem it is convenient to use
states  $\langle \hat{B}(p) | \hat{B}(q) \rangle  =  (2\pi)^3  \de^{(3)}(\vec{p} -  \vec{q})$ normalised in a non-relativistic manner (the translation to the usual  states is $\state{\hat{B}} =   \state{{B}}/\sqrt{2 E_B}$). 
Taking the derivative of $ \matel{\hat{B}}{H}{\hat{B}}$ (using 
$ \partial_{m_q} \langle \hat{B}(p) | \hat{B}(q) \rangle =0 $) one obtains
\begin{equation}
m_q \partial_{m_q} E_B = m_q \matel{\hat{B}}{\bar qq}{\hat{B}} \;,
\end{equation}
which is equivalent to 
\begin{equation}
m_q \partial_{m_q} 2 E_B^2 = 2  m_q \matel{{B}}{\bar qq}{{B}} \;,
\end{equation}
which in turn is consistent with 
\begin{equation}
 m_B^2|_{m_q} = \sum_q  m_q  \matel{{B}}{\bar qq}{{B}} \;,
\end{equation}
since the momenta are independent of the mass.  This is the relation quoted in \eqref{eq:FH} 
in the main text.

\subsection{$\Delta m_\pi|_{\text{QED}}$ from soft theorem and Weinberg sum rules} 
\label{app:DeltaPi}

Using soft-pion techniques it was shown that \cite{Das:1967it} 
\begin{equation}
\label{eq:CA}
\Delta m_\pi|_{\text{QED}}= \frac{ 3\al}{ 8 \pi m_\pi f_\pi^2} \int_0^\infty ds s \ln\frac{\mu^2}{s} (\rho_V(s) - \rho_A(s)) +  \ORD(m_\pi^2/m_\rho^2) \;,
\end{equation}
where $\rho_{V} =  f_\rho \de(s-m_\rho^2) + \dots$ is the spectral density of the vector triplet current and $\rho_A$ is the analogous quantity for the axial case.  The $\ln s$-term originates from integrating over the photon momentum $d^4 q$. 
We refer the reader to \cite{Donoghue:1996zn}  
for an improved treatment using chiral perturbation theory.  In fact, as is the case 
for all soft-pion results,   \EQ\eqref{eq:CA} follows from the LO electromagnetic term in the Lagrangian and can therefore be systematically improved beyond the soft limit
to the extent that its low energy constants (i.e. couplings) are known.
Using the Weinberg sum rules \cite{WSR}, which are phenomenologically successful,  
a good estimate was obtained \cite{Das:1967it}. Taking the equations resulting from the so-called first and second Weinberg sum rule in  \cite{Zwicky:2016lka}, then
\begin{equation}
f_\rho^2 = f_{a_1}^2 + f_\pi^2\;, \quad   m_\rho^2 f_\rho^2 = m_{a_1}^2 f_{a_1}^2 \;,
\end{equation}
(where the chiral limit $m_q =0$ is assumed). Moreover, the spectral functions are truncated after the first vector
meson resonances $\rho$ and $a_1$ which can be justified as the chiral symmetry is restored at high energy.
Using these in expressions in \eqref{eq:CA} one gets 
\begin{equation}
\label{eq:CA2}
\Delta m_\pi|_{\text{QED}}= \frac{ 3\al }{8 \pi} \frac{m_\rho^2 f_\rho^2}{ m^2_\pi f_\pi^2} m_\pi  \ln \frac{f_\rho^2}{f_\rho^2 - f_\pi^2} \approx 4.8 \MeV \;,
\end{equation}
for $f_\pi = 131\MeV$, $m_\rho = 0.77\MeV$ \cite{PDG} and $f_\rho = 215 \MeV$ \cite{Bharucha:2015bzk}. 
Since the quark mass effect is small  $\ORD( (m_u-m_d)^2)$ \eqref{eq:PiKaon}, one has 
$\Delta m_\pi \approx \Delta m_\pi|_{\text{QED}}$ which is rather close to the experimental value  
$\Delta m_\pi \ = +4.5936(5)\,\MeV$ \cite{PDG}.  Clearly \eqref{eq:CA2} is a crude approximation 
as  more detailed analyses \cite{Gross:1979ur,Donoghue:1996zn} including finite width effects yields a result which is ca $+ 1.2 \MeV$ larger  \cite{Gross:1979ur}. We therefore assign an uncertainty of this amount to 
$\Delta m_\pi|_{\text{QED}}$ in \TAB\ref{tab:overview}.

It is also worthwhile to mention two other interesting aspects in conjunction with $\Delta m_\pi|_{\text{QED}}$.
First, by using by using QCD inequalities it has been shown that 
$\Delta m_\pi|_{\text{QED}} \geq 0$ \cite{Witten:1983ut} which is of course well satisfied.
Second Dashen's theorem \cite{Dashen:1969eg} states that  
$\Delta m_\pi^2|_{\text{QED}} - \Delta m_K^2|_{\text{QED}}  = \ORD(\al m_s,\al m_q \ln m_q)$ 
as a result of degeneracy in the $SU(3)_F$ limit  $m_s = m_d = m_u$.
The corrections seem rather large and are largely kinematic, the larger $K$ mass in the Kaon propagator 
\cite{Donoghue:1993hj}. 
Lattice Monte Carlo simulations  have settled this matter to large precision  \cite{Fodor:2016bgu} 
(cf.\,\cite{Portelli:2015wna} for a review).

\bibliographystyle{utphys}
\bibliography{../../Refs-dropbox/References_QED.bib}

\end{document}